# Giant magnetic anisotropy energy and long coherence time of uranium substitution on defected $Al_2O_3(0001)$

Jie Li, Lei Gu and Ruqian Wu*

*Department of Physics and Astronomy, University of California, Irvine, California 92697-4575, USA.*

Nanomagnets with giant magnetic anisotropy energy and long coherence time are desired for various technological innovations such as quantum information procession and storage. Based on the first-principles calculations and model analyses, we demonstrate that a single uranium atom substituting Al on the $Al_2O_3(0001)$ surface may have high structural stability and large magnetic anisotropy energy up to 48 meV per uranium atom. As the magnetization resides in the localized f-shell and is not much involved in chemical bonding with neighbors, long coherence time up to ~1.6 $mS$ can be achieved for the quantum spin states. These results suggest a new strategy for the search of ultrasmall magnetic units for diverse applications in the quantum information era.

*Author to whom correspondence should be addressed. Electronic mail: wur@uci.edu.

# I. INTRODUCTION

As potential building blocks for quantum computing [1-3] and data storage [4-5] devices, nanomagnets and molecular magnets have received increasing attention in recent years. Exceedingly large magnetic anisotropy energy (MAE) is typically required to combat with thermal fluctuation and to set appropriate energy levels for quantum operations. However, MAEs of most nanomagnets are only a few tenth millielectronvolts (meV), that vitally inhibits their exploitation. Therefore, it is desired to search for new magnetic systems with large MAEs and, furthermore, long coherence time of their quantum spin states. So far, several systems were reported to have MAE > 30 meV, such as freestanding and supported transition metal dimers [6-9] or single transition metal adatoms on CuN, Rh and MgO(100) surfaces [10-13]. It is recognized that one needs to place magnetic atoms with strong spin-orbit coupling (SOC) in an environment with weak crystal field for attaining large MAE. Furthermore, the reduction of the spin-vibration coupling is another important issue for magnetic units being used in quantum information devices.

Rare earth and actinide atoms have unique advantages as their magnetism stems from the localized *f*-shells, which are protected by the *s, p* and *d* outer shells and are not much involved in chemical bonds with their neighbors in compounds or alloys. Molecular magnets with rare earth and actinide ingredients have been extensively investigated in different research fields such as for manipulating of nuclear spins [14, 15], attaining quantized spin states [16], and achieving atomic clock transitions for robust qubits [17]. With strong SOC and small crystal field splitting, the behavior of *f*-electrons is not much different from that in isolated atoms, i.e., with large orbital moments and quantized energy levels with *j* being a good quantum number. As their magnetic features are not directly related to chemical bonding with neighbors, the spin-vibration coupling is expected to be weak, suitable for getting long coherence time of entangled quantum spin states. Ideally, systems with rare earth and actinide magnetic atoms may become promising candidates for quantum information processing and storage purposes.

In this work, we investigate the electronic and magnetic properties of a single uranium atom as a substituent on the $Al_2O_3(0001)$ surface ($U/Al_2O_3$) through density functional theory (DFT) calculations and electrostatic model analyses. The similar trivalent characteristics between uranium and aluminum atoms ensure the least disturbance to the local electronic balance and the insulating state of the film. This system is found to have a magnetic anisotropy energy as large as 48 meV per U atom and a long relaxation time ($1.6\ mS$) at a reasonably high temperature (10K), which may offer strong thermal stability as needed for practical use. Results of vibrational spectra and diffusion energies also suggest the high structural stability. This research suggests a new strategy for the design of emergent quantum materials.

## II. METHODOLOGY

As depicted in Fig. 1, a 2×2 supercell in the lateral plane, with 18 layers of atoms and a vacuum of 15Å thick along the surface normal, was used to mimic the periodic sapphire $Al_2O_3(0001)$ surface. The lattice constant in the lateral plane was fixed according to the experimental value of the bulk α-$Al_2O_3$ (a = b = 4.76 Å). Aluminum atoms in two surface layers were substituted by uranium atoms to keep the inversion symmetry for the computational convenience. This corresponds to a surface density of uranium substitution at $1.27\times10^{18}$ m$^{-2}$. DFT calculations were carried out using the Vienna ab-initio simulation package (VASP), at the level of the spin-polarized generalized-gradient approximation (GGA) with the functional developed by Perdew-Burke-Ernzerhof (PBE) [18]. Hubbard U was adopted to describe the electron correlation for the *f*-electrons of uranium, with typical values of U = 4.50 eV and J = 0.54 eV [19]. The interaction between valence electrons and ionic cores was considered within the framework of the projector augmented wave (PAW) method [20-21]. The energy cutoff for the plane wave basis expansion was set to 500 eV. To sample the two-dimensional Brillouin zone, we used a Gamma-centered 8×8 k-grid mesh. All atoms were fully relaxed using the conjugated gradient method for the energy minimization until the force on each atom became smaller than 0.01 eV/Å, and the energy convergence in all DFT calculations in this work was better than $10^{-8}$ eV.

## III. RESULTS AND DISCUSSION

We first determined the binding energies of uranium atoms according to:

$$\Delta E_{U/Al_2O_3} = E_{Al_2O_3} + \mu_U - E_{U/Al_2O_3} \qquad (1)$$

where $E_{Al_2O_3}$ and $E_{U/Al_2O_3}$ represent the total energies of the defected $Al_2O_3$(0001) surface without and with uranium taking the vacancy site, respectively. $\mu_U$ is the chemical potential of a uranium atom which was set to be equal to the energy per atom in a uranium dimer. After the removal of the topmost Al atoms, U takes the Al-vacancy ($Al_V$) site with a huge energy gain, up to $\Delta E_{U/Al_2O_3}$ =14.9 eV. This is even larger than the corresponding binding energy of Al atom (12.1 eV) and clearly indicates that uranium atoms are strongly anchored to the defected $Al_2O_3$(0001) surface. As uranium atom has a bigger size than aluminum atom, it shifts out of the surface plane by 1.24 Å and the bond length between uranium and the nearest three oxygen atoms is 2.14 Å.

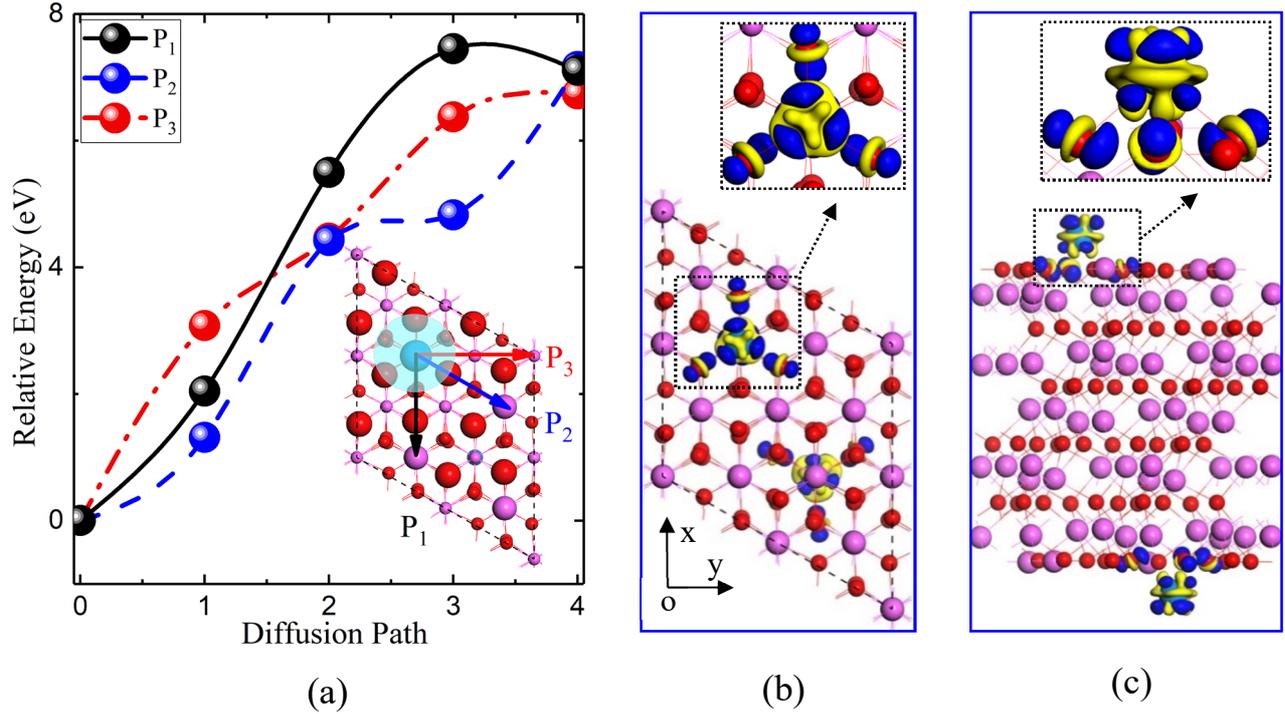

Figure 1. (a) Three diffusion pathways of uranium atom (inset) and corresponding energy profiles. Red and pink spheres represent O and Al atoms, respectively. To make the surface O and Al atoms more distinguishable, they are represented by larger spheres. The arrows represent the diffusion pathways for energy calculations. (b) (c) The top and side views of charge density difference, i.e., $\rho(U/Al_2O_3)- \rho(Al_2O_3)- \rho(U)$. Blue and yellow colors represent charge accumulation and depletion, respectively.

We further considered the structural stability of U/Al$_2$O$_3$ against sideway displacements of uranium along pathways depicted as P$_1$, P$_2$, and P$_3$ in Fig. 1(a). While doing so, we fixed the (x,y) coordinates of the uranium atom along the pathways and allowed all other atoms as well as the z-coordinate of the uranium atom to fully relax. From the relative energies of diffusion, one may see that uranium atom is tightly anchored at the Al$_V$ site, with rapid increases of energy as it drifts away along all three pathways. In the other word, uranium atoms may quickly segregate to the substitutional sites on the Al$_2$O$_3$(0001) surface in experiments since the binding energy for them at the Al$_v$ site is significantly lower than those on the flat regions, which offers a large possibility to deposit single uranium atoms on the defected $Al_2O_3$ (0001) surface.

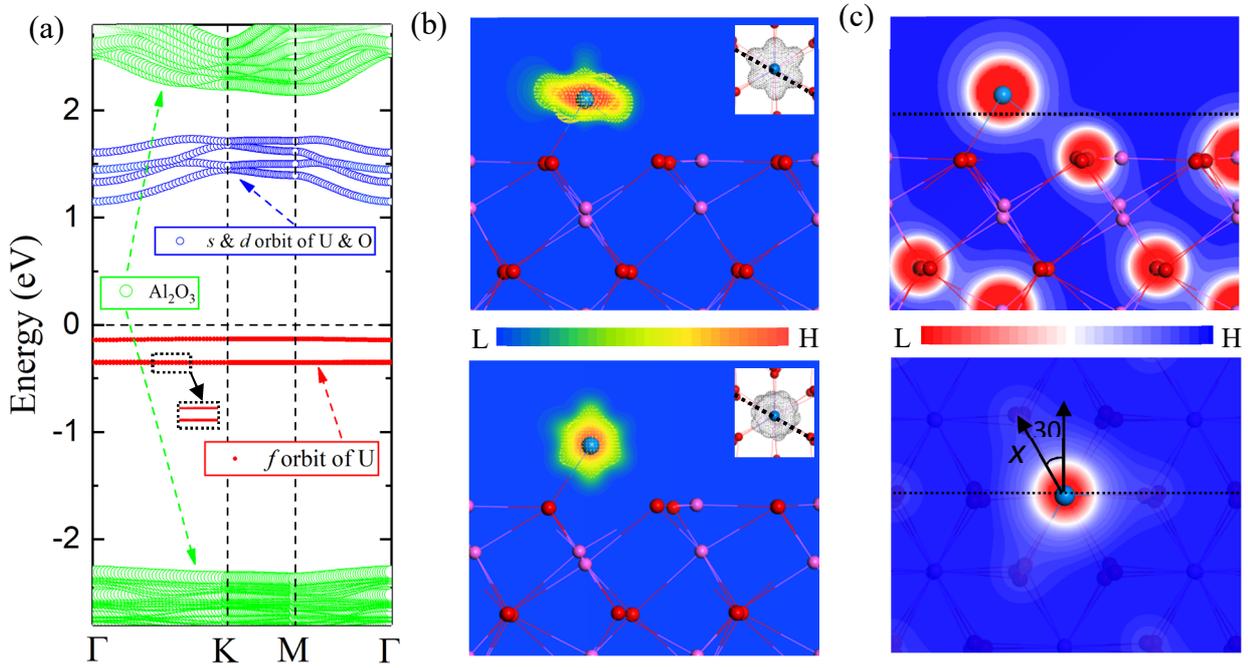

*Figure 2. (a) The band structure of $U/Al_2O_3$. (b) Charge density of f-electrons of $U/Al_2O_3$ with the magnetization axis pointing along the normal and in-plane direction, respectively. (c) Side and top views of electrostatic potential of $Al/Al_2O_3$ with Al atoms at the U site.*

Each trivalent uranium atom donates three electrons to its oxygen neighbors, as shown by the charge density difference in Figs. 1 (b) and (c). The charge transfer is limited to the surface layer, and a slight surface reconstruction occurs. The band structure of $U/Al_2O_3$ in Fig. 2(a) shows that the s and d bands of U are empty, and three flat f bands of U are occupied in the wide band gap of $Al_2O_3$(0001). This is ideal as we seek for the separation of sources of chemical bonding (the outer s and d electrons) and magnetization (the localized f-electrons). Furthermore, the insulating state with large band gap is beneficial for resisting surface oxidation and minimizing the coupling between spin and electron excitations. The calculated spin magnetic moment of each uranium atom is ~3 $\mu_B$, accompanied by small negative magnetic moments of ~ 0.09 $\mu_B$ from three neighboring O atoms. Due to the strong localization and spin-orbit coupling in the f-shell, each uranium atom also has a large orbital magnetic moment up to 3.2 $\mu_B$ in antiparallel to the spin moment, very different from cases with transition metal magnetic atoms for which the orbital magnetic moments are mostly quenched.

The U atom embedded in $Al_2O_3$(0001) has a $C_{3v}$ local symmetry and the standard spin Hamiltonian describing the single-ion anisotropy can be written as

$$H_{spin} = D_{xx}S_x^2 + D_{yy}S_y^2 + D_{zz}S_z^2 + D_{xy}(S_xS_y + S_yS_x)$$

$$+ D_{xz}(S_xS_z + S_zS_x) + D_{yz}(S_yS_z + S_zS_y) \qquad (2)$$

where D-tensor represents the magnetic anisotropy parameters. According to the four states mapping method [22], we calculated total energies of different spin configurations with the SOC term in the self-consistence (details are shown in the supporting Information). By fitting these energies, the D-tensor are obtained, as shown in Table. SI. In particular, the perpendicular and in-plane magnetic anisotropy energies are given by $D_{xx}S_x^2 - D_{zz}S_z^2$ (-48.1 meV) and $D_{xx}S_x^2 - D_{yy}S_y^2$ (-26.0 meV), which indicate that this system has an in-plane easy axis along the x direction. Large anisotropy energy barriers for the magnetic moment rotating toward both y and z axes suggest that the magnetization is strongly pined along the x axis at a reasonably high temperature. To ensure the reliability of these MAEs, test calculations with denser k-point meshes (from 16 to 100 points in the Brillouin zone) and different values of Hubbard U (0-6eV) were done (see Fig. S1 and Fig. S2 in the supporting Information). All results show that MAEs are converged in the present calculations.

To understand the mechanism of the giant MAEs of U/$Al_2O_3$, we used a simple electrostatic model which has been applied for analyzing the preferential direction of magnetization in $f$-electron materials [16, 23]. The angle dependent energies in Eq. (2) can be estimated from the local crystal field $V_{CF}(\theta,\varphi)$ and the Sievers charge density of uranium atom $\rho_{(\theta_S,\varphi_S)}(\theta,\varphi)$

$$E(\theta_S,\varphi_S) = \int_{\theta=0}^{\pi}\int_{\varphi=0}^{2\pi} V_{CF}(\theta,\varphi)\rho_{(\theta_S,\varphi_S)}(\theta,\varphi)\sin(\theta)\,d\theta d\varphi \qquad (3)$$

Here, $\rho_{(\theta_S,\varphi_S)}(\theta,\varphi)$ is the charge density of $f$-electrons that follows the spin orientation $(\theta_S,\varphi_S)$ as depicted in Fig. 2(b), and $V_{CF}(\theta,\varphi)$ is extracted from the electrostatic potential which includes the ionic, Hartree and exchange correlation parts as shown in Fig. 2(c). For U/Al$_2$O$_3$(0001), three oxygen anions have the most important influence on the uranium atom and $V_{CF}(\theta,\varphi)$ has a $C_{3v}$ symmetry. Considering the two possible orientations of the easy axis, i.e., normal (the z axis) and in-plane directions (the x axis), we plotted the corresponding charge densities of the $f$-electrons of Uranium in Fig. 2(b). One may see that the distribution pattern of $5f$ electrons somewhat follows the rotation of magnetization. Because $5f$-orbitals are rather large, we may divide them into two parts, i.e., localized part within a small sphere that rotates with spin moment due to the strong SOC, and spilled out part a distance away from uranium nuclei that are controlled by the crystal field [24]. The overall spatial distribution of $f$ electrons of uranium is more complicated than either $4f$ electrons of rare earth atoms or 3d electrons of transition metal atoms [25]. When the spin moment is set along the z-axis, the $f$-electrons of uranium form a bowl-shaped distribution as shown at the top of Fig. 2(b). Although $m$ is no longer a good quantum number under the influence of the crystal potential, we see that $m=\pm 3$ components are well separated from others as seen from the $m$-projected density of states in Fig. 3(a) as the spin is align along the z-axis, a case $\rho_{(\theta_S,\varphi_S)}(\theta,\varphi)$ also has the C$_{3v}$ symmetry. Nevertheless, the weight in $m=-3$ is much larger than that in $m=+3$, showing the orbital magnetization. As the lowest non-spherical term in the crystal field is $Y_{3,\pm 3}$, $m=\pm 2$ and $m=\pm 1$ parts are degenerated. When the spin turns to the x-axis, the symmetry of $\rho_{(\theta_S,\varphi_S)}(\theta,\varphi)$ is lowered. All m-components intermix as we projected the f-orbitals to spherical harmonics with a local "z"-axis (actually the x-axis as shown in the inset of Fig. 3(b)) following the rotation of spin. As the lobs of *f*-orbitals pointing toward O atoms, some $Y_{3,-3}$ and $Y_{3,\pm 1}$ electron transfer to $Y_{3,3}$ and $Y_{3,0}$ due to influence of crystal field, and the spatial distribution for *f* electrons is more isotropic as shown at the bottom of Fig. 2(b). This is also reflected in the PDOS of *f*-orbits of uranium atom in Fig. 3(b): three *f*-electrons almost evenly distribute in seven *f* orbits. It is obvious that the electric potential in Fig. 2(c) has significant non-spherical components around the Al or $Al_V$ site. The large MAE of $U/Al_2O_3$ is not only directly from SOC as for most transition metals, the crystal field also contributes due to the significant redistribution of *f*-electrons during spin reorientation.

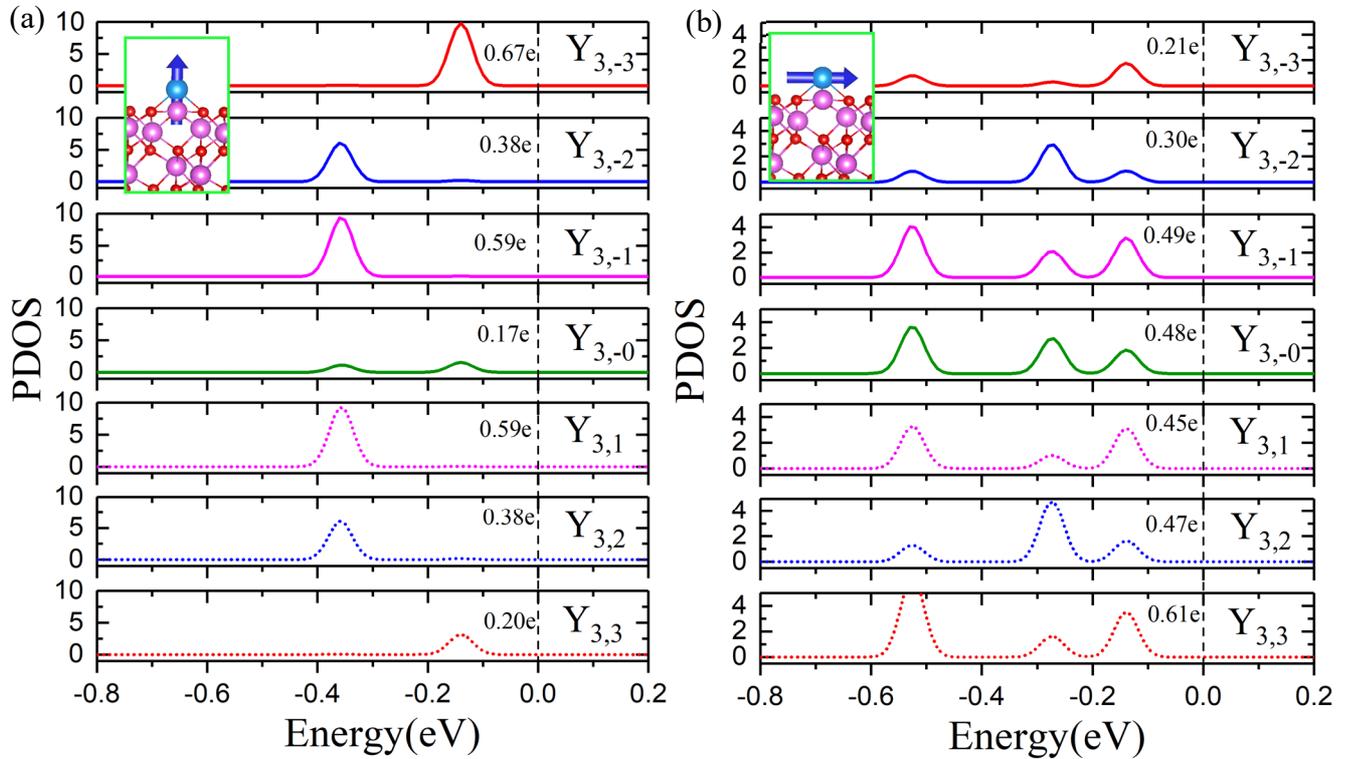

Figure 3. (a) and (b) The projected density of states (PDOS) of f-orbitals of uranium atom with the quantization axis along normal direction and in the plane, respectively.

From the viewpoint of quantum computing, it is very important to have the magnetic sources being protected against the lattice vibration which plays an important role in magnetic fluctuation and magnonic dissipation. To this end, we explore the spin-vibration coupling (SVC) as shown in Fig. 4. From our DFT calculations, the effect of spin reorientation on the phonon spectrum is small but obviously visible, especially for the low frequency bands that come from the motion of uranium atom. It appears that vibrations shift to lower frequency when spin aligns in the lateral plan as shown in Fig. 4(a). For example, vibration energies of three uranium modes at the Γ point (7.14 meV, 9.29 meV and 11.77 meV) shift by -0.08 meV, 0.51 meV and 0.26 meV as we rotate spin from the x- to z-axis. As a step to estimate the SVC strength, we explored the effect of these vibrations on MAE by displacing the uranium (0.2 Å) and surrounding O and Al atoms according to the displacement vectors of these normal modes shown in the Fig. 4(a). The corresponding MAEs are 50.8, 57.5 and 88.7 (75.5, 57.5 and 67.1) meV for the case of uranium atoms moving along (reverse along) the three vibrational vectors, respectively. The mutual influence between vibration and spin reorientation of uranium suggests strong SVC in this system and hence the energy exchange across the two excitations is still a concern for the system hosting quantum information.

As the D-tensor are computed from energies ($E_i$) of four different spin configurations as

$$D_{ij} = \frac{1}{4S^2}(E_1 + E_4 - E_2 - E_3) \qquad (4)$$

we may further obtain the spin-vibration coupling coefficients by taking the derivatives of $E_i$ with respect to $u_{ka}$ (the displacements of atom k along $a$ direction), i.e.

$$\frac{\partial D_{ij}}{\partial u_{ka}} = \frac{1}{4S^2}\left(\frac{\partial E_1}{\partial u_{ka}} + \frac{\partial E_4}{\partial u_{ka}} - \frac{\partial E_2}{\partial u_{ka}} - \frac{\partial E_3}{\partial u_{ka}}\right) \qquad (5)$$

Here, $-\frac{\partial E_i}{\partial u_{ka}}$ ($i$=1,…,4) is the force acting on the atom $k$ along the $a$ direction, which can be connected to the Hellmann-Feynman force in DFT schemes with the plane wave bases. In this work, we considered motions of the uranium atom and its three O neighbors. For U/Al$_2$O$_3$ (a spin 3/2 system), the $S_{x,y,z}$ in Eq. (2) are 4x4 matrices. Using DFT parameters, there are 4 quantum spin states or two Kramers doublets for U/Al$_2$O$_3$ as shown in Fig. 4(b). Due to the time reversal symmetry, phonon induced transition within each doublet is forbidden [26]. Therefore, the magnetic relaxation pathways for a ground state are those illustrated in Fig. 4(b), (the detail as shown in the supporting Information). The corresponding relaxation times of these quantum spin states were calculated by using the Non-equilibrium Green's function method and the master equation. Following as Markov dynamics, the evolution of the system can be described by

$$\frac{d}{dt}p_{S_i}(t) = \sum_{S_j}\left[\gamma_{S_j}^{S_i}p_{S_j}(t) - \gamma_{S_i}^{S_j}p_{S_i}(t)\right] \quad (6)$$

where $p_{S_i}$ and $\gamma_{S_j}^{S_i}$ represent the probability of being at spin state $S_i$ and the transition rate from spin state $S_j$ to spin state $S_i$, respectively. As the energy barrier of the spin states is accessible to the phonon bath (0~120meV, as shown in Fig. S3 in the supporting Information) in this work, we only considered the relaxation caused by the first order spin-phonon coupling, i.e., single phonon mode. Then, the transition rate from spin state $S_i$ to spin state $S_j$ is given by:

$$\gamma_{S_i}^{S_j} = \sum_q \frac{i}{\hbar} G_q^<(\omega)|a_q|^2 \quad \text{and}$$

$$G_q^<(\omega) = -\frac{i\pi}{\omega_q}\{\delta(\omega - \omega_q)N(\omega_q) + \delta(\omega + \omega_q)[N(\omega_q) + 1]\} \quad (7)$$

where $G_q^<(\omega)$ is the lesser Green's function for phonon q, and $a_q$ is the transition matrix element of SVC, $a_q = \left\langle S_i \left| \frac{\partial H_{spin}}{\partial V_q} \right| S_j \right\rangle$ with $V_q$ represent the atomic displacements corresponding to phonon mode q [27] (the detail calculations are shown in the supporting Information). By substituting

$$p_{S_j}(t) = \sum_k \varphi_{S_j}^{(k)}\exp(-t/\tau_k) \quad (8)$$

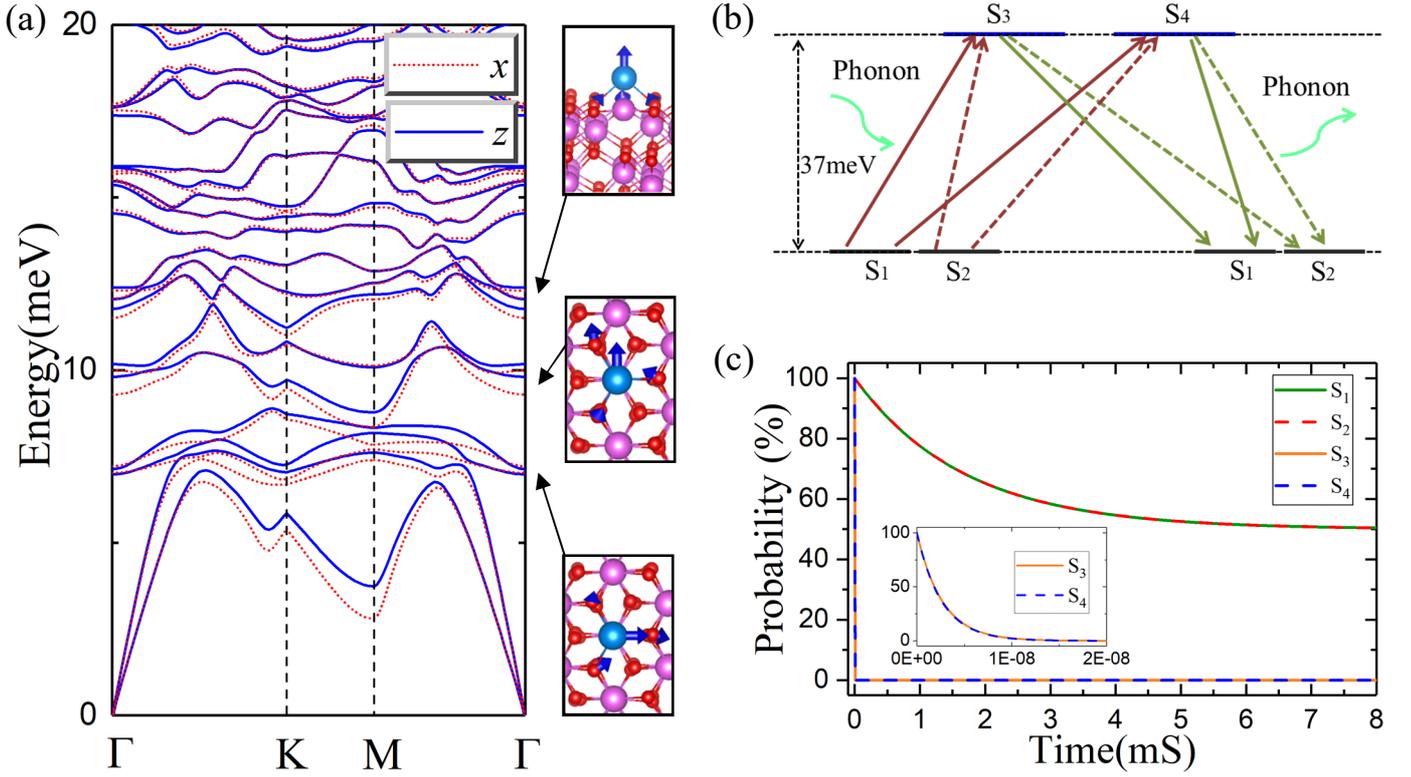

*Figure 4. (a) The phonon spectrums of U/Al$_2$O$_3$ and the schematic moving of uranium atom and three nearest oxygen atoms in three vibration models (high lines) which mostly come from the contributions of Uranium atom. (b) The schematic magnetic relaxation pathways of spin states. (c) The decaying of spin states with time.*

into the Eq. (6), we have

$$-\frac{1}{\tau_k}\varphi_{S_j}^{(k)} = \sum_{S_i}\left[\gamma_{S_i}^{S_j}\varphi_{S_i}^{(k)} - \gamma_{S_j}^{S_i}\varphi_{S_j}^{(k)}\right] = \sum_{S_i}\left[\gamma_{S_i}^{S_j} - \delta_{S_i}^{S_j}\sum_{S_l}\gamma_{S_l}^{S_i}\right]\varphi_{S_i}^{(k)} \quad (9)$$

Therefore, eigenvalues of the coefficient matrix

$$\Gamma_{S_i}^{S_j} = \gamma_{S_i}^{S_j} - \delta_{S_i}^{S_j}\sum_{S_l}\gamma_{S_l}^{S_i} \quad (10)$$

give the time scales. There is a zero value corresponding to the statistic state, and the smallest nonzero value characterizes the relaxation of magnetization [28]. The probabilities of holding in these spin states through relaxation are shown in Fig. 4 (c). While the excited doublet quickly decays (within *nS* as shown in the inset), the ground states decay rather slowly, due to the high energy barrier raised by the magnetic anisotropy. We may see from Fig. 4 (c) that a long coherence time (1.6 $mS$) at a reasonably high temperature (10K) for the $S_1$ and $S_2$ states. This implies that U/$Al_2O_3$ can be a potential candidate for being developed as a quantum material.

In end, we also suggest that if the substituents are changed to rare earth atoms, such as Praseodymium and Erbium atoms, the corresponding MAE can be even higher than Uranium. As the SVC in rare earth system is much weaker than uranium as well, longer relaxation time can be expected, which may deserve experimental and theoretical explorations in the future.

## IV. CONCLUSION

In summary, we investigated the structural and electronic properties of a single uranium atom as a substituent on the $Al_2O_3$ (0001) surface. The large binding energy and high diffusion energy barriers suggest that the uranium atoms are strongly anchored to the substitutional sites. The MAE of this spin 3/2 system is as large as 48 meV and the relaxation time is up to ~1.6 $mS$ at 10K. Thus, we recommend this system or similar structures with rare earth substituents as a new class of potential candidates for developing materials for quantum information processing and storage.

## Acknowledgments

Work was supported by the Department of Energy (grant No. DE-SC0019448). DFT calculations were performed on supercomputers at NERSC.